\documentclass[prl,twocolumn,showpacs,preprintnumbers,
amsmath,amssymb,tightenlines,epsfig]{revtex4}

\usepackage{bm}% bold math
\usepackage{graphicx}% Include figure files

\newcommand{\ej}{\epsilon_j}
\newcommand{\beq}{\begin{equation}}
\newcommand{\eeq}{\end{equation}}
\newcommand{\bea}{\begin{eqnarray}}
\newcommand{\eea}{\end{eqnarray}}

\newcommand{\<}{\langle}
\renewcommand{\>}{\rangle}
\renewcommand{\(}{\left(}
\renewcommand{\)}{\right)}
\renewcommand{\[}{\left[}
\renewcommand{\]}{\right]}
\newcommand{\commentout}[1]{{}}
\newcommand{\half}{\hbox{$1\over2$}}

\renewcommand{\d}{\dag}
\newcommand{\h}{\hat}
\newcommand{\p}{\partial}
\newcommand{\f}{\frac}
\newcommand{\s}{\sum}

\newcommand{\rro}{\right)}
\newcommand{\lro}{\left( }

\begin{document}

\title{Nonadiabatic dynamics of a Bose-Einstein condensate in an optical lattice}
\author{Lorenzo Isella}
\author{Janne Ruostekoski}
\affiliation{Department of Physics, Astronomy and Mathematics, University of
Hertfordshire, Hatfield, Herts, AL10 9AB, UK}
\begin{abstract}
We study the nonequilibrium dynamics of a Bose-Einstein condensate
that is split in a harmonic trap by turning up a periodic optical
lattice potential. We evaluate the dynamical evolution of the phase
coherence along the lattice and the number fluctuations in
individual lattice sites within the stochastic truncated Wigner
approximation when several atoms occupy each site. We show that the
saturation of the number squeezing at high lattice strengths, which
was observed in recent experiments by Orzel {\it et.\ al.}, can be
explained by the nonadiabaticity of the splitting.
\end{abstract}

\date{\today}
\maketitle

Ultra-cold atomic gases in periodic optical lattice potentials have
recently attracted considerable interest and inspired experiments,
e.g., on the Bose-Einstein condensate (BEC) coherence
\cite{AND98,HAD04}, superfluid dynamics \cite{CAT01,FOR03,MOR01},
the number-squeezed \cite{ORZ01} and the Mott insulator (MI)
\cite{GRE02a} states, and quantum information applications
\cite{PEI03,MAN03}. An optical lattice provides a clean
many-particle system with enhanced interactions, resulting in a
unique opportunity to study strong quantum fluctuations. While the
classical mean-field theories, such as the Gross-Pitaevskii equation
(GPE), have been successful in describing the full multi-mode
dynamics of weakly-interacting BECs, they have severe limitations in
optical lattices, as they disregard thermal and quantum
fluctuations, decoherence, and the information about quantum
statistics. In this paper we study matter wave dynamics beyond the
GPE by considering a harmonically trapped finite-temperature BEC
that is dynamically split by an optical lattice potential. We show
that the experimentally observed saturation of the number squeezing
at high lattice strengths \cite{ORZ01} can be explained by the
nonadiabaticity of the loading of atoms into the lattice. The
thermal and quantum fluctuations are included within the truncated
Wigner approximation (TWA). The multi-mode TWA provides a natural
representation for the dynamical fragmentation of the initially
uniform BEC in the lattice and the transition to the regime that can
also be described by the discrete Bose-Hubbard Hamiltonian (BHH).
Moreover, the resulting highly occupied number squeezed states are
also of great interest in the Heisenberg limited interferometry
\cite{HOL93,ORZ01}.

We study the loading of atoms into the lattice within the TWA. The
TWA may be obtained by using the familiar techniques of quantum
optics \cite{GAR,DRU93} to derive a generalized Fokker-Planck
equation (FPE) for the Wigner distribution of the trapped multi-mode
BEC \cite{Steel}. The TWA consists of neglecting the dynamical
quantum noise, acting via third-order derivatives in the FPE, and
results in a deterministic equation for the classical field $\psi_W$
which coincides with the GPE:
\beq
\label{GP}
i\p_t\psi_W=\mathcal{L}\psi_W+g|\psi_W|^2\psi_W \,,
\eeq
where ${\cal L}\equiv -\hbar^2\nabla^2/(2m)+V$. The thermal and
quantum fluctuations are included in the initial state of $\psi_W$
in Eq.~(\ref{GP}) which represents an ensemble of Wigner distributed
wave functions. The neglected terms are small when the amplitudes of
the Wigner distribution are large. The TWA and closely related
approaches have previously been successful in describing atomic BECs
\cite{Steel,SIN01,SIN02,POL03,POL03b,DAV02} and optical squeezing
\cite{COR01}. In particular, the TWA is shown to produce correctly,
e.g., the Beliaev-Landau damping \cite{SIN02} and it has been argued
that the TWA more generally provides an accurate description for the
short-time asymptotic behavior of the full quantum dynamics
\cite{POL03b}.

We consider a BEC in a tight elongated cigar-shaped trap, with the
trap frequencies $\omega\equiv\omega_x\ll
\omega_{y,z}\equiv\omega_\perp$, and ignore the density fluctuations
along the transverse directions. This results in an effective 1D GPE
for $\psi_W(x,t)$ with $g=g_{\rm 1D}=2\hbar\omega_\perp a$ in
Eq.~(\ref{GP}), where $a$ denotes the scattering length. The BEC is
initially assumed to be in thermal equilibrium in a harmonic trap
$V_h(x)=m\omega^2 x^2/2$. A self-consistent calculation of the
initial state would involve solving the coupled
Hartree-Fock-Bogoliubov equations for the condensate and
non-condensate populations \cite{HUT97}. Here we resort to a simpler
Bogoliubov approximation and expand the field operator
$\h\psi(x,t=0)$ in terms of the BEC ground state amplitude
$\h\alpha_0\psi_0$, with $\<\h\alpha_0^\dagger\h\alpha_0\>=N_0$, and
the excited states:
\beq
\label{field} \h\psi(x)=\psi_0(x)\h\alpha_0+ \sum_{j>0} \big[
u_j(x)\h\alpha_j-v^*_j(x)\h\alpha^\d_j \big]\,,
\eeq
where $u_j(x)$ and $v_j(x)$ ($j>0$) are obtained from
\begin{align}
\label{Bogo}
 \lro \mathcal{L}-\mu+2N_0g_{1D}|\psi_0|^2\rro u_j-N_0g_{1D}\psi_0^2 v_j    & = \ej u_j,\nonumber\\
\lro \mathcal{L}-\mu+2N_0g_{1D}|\psi_0|^2\rro v_j-N_0g_{1D}\psi_0^{*2} u_j & =-\ej v_j\,.
\end{align}
Here $\h\alpha_j$ are the quasiparticle annihilation operators, with
$\<\h\alpha_j^\dagger \h\alpha_j\>=\bar{n}_j\equiv [\exp{(\beta
\ej)}-1]^{-1}$, $\beta\equiv1/k_BT$, and $\psi_0$ is ground state
solution of the GPE with the chemical potential $\mu$.

In the Wigner description we replace the quantum operators
$(\h\alpha_j,\h\alpha_j^\d)$ (for $j>0$) by the complex random
variables $(\alpha_j,\alpha_j^*)$, obtained by sampling the
corresponding Wigner distribution of ideal harmonic oscillators in a
thermal bath \cite{GAR}:
\beq
\label{wigner}
W(\alpha_j,\alpha_j^*)=\f{2}{\pi}\tanh \( \xi_j\)
\exp\[ -2|\alpha_j|^2\tanh\( \xi_j\)\]\,,
\eeq
where $\xi_j\equiv \beta\ej /2$. The Wigner function is Gaussian
distributed with the width $\bar{n}_j+\half$. The nonvanishing
contribution to the width at $T=0$ for each mode represents the
quantum noise. The Wigner function returns symmetrically ordered
expectation values, so $\<\alpha_j^*\alpha_j\>_W = \bar{n}_j+\half$,
and $\<\alpha_j\>_W=\<\alpha_j^*\>_W=\<\alpha_j^2\>_W=0$, etc.

For large BECs, $N_0\gg1$, the main contribution to the matter wave
coherence in the superfluid regime is due to the thermal and quantum
fluctuations of low-energy phonons and the quantum fluctuations of
the initial harmonically trapped BEC mode are not very important.
Consequently, we could treat the BEC mode $\h\alpha_0$ even
classically. However, here we assume it to be in a coherent state
and sample the quantum fluctuations according to the corresponding
Wigner distribution \cite{GAR}:
$W(\alpha_0,\alpha_0^*)=2\exp{(-2|\alpha_0 - N_0^{1/2}|^2)}/\pi$, so
that $\<\alpha_0\>_W=N_0^{1/2}$ and $\<\alpha_0^*\alpha_0\>_W
=N_0+\half$. Since we compare the matter wave coherence between the
atoms in different lattice sites, the global BEC phase is
unimportant. The advantage of using the coherent state description
is that the Wigner function is positive.

Due to the symmetric ordering of the expectation values obtained
from the Wigner distribution, it is difficult, or even impossible,
to extract several correlation functions for the full multi-mode
field operator, since the Wigner field is symmetrically ordered with
respect to {\it every} mode. In \cite{Steel} the phase diffusion of
a BEC was therefore calculated by defining a `condensate mode'
operator associated with the projection of the stochastic field onto
the ground state solution. Since we study the splitting of a BEC by
a periodic optical lattice potential, it is useful to define
analogously the ground state operators $a_j$ for each individual
lattice site $j$:
\beq
\label{projection}
a_j(t)=\int_{j^{\rm th} {\rm well}}dx\,
\psi^*_{0}(x,t)\psi_W(x,t)\,,
\eeq
where $\psi_W(x,t)$ is the stochastic field, determined by
Eq.~(\ref{GP}), and $\psi_0(x,t)$ is the ground state wave function
at time $t$, obtained by integrating the GPE in imaginary time in
the potential $V(x,t)$. The integration is over one lattice site.
For each lattice site ground state mode $a_j$, the normally ordered
expectation values can be easily obtained $\< a_i^*a_j\>_W= \langle
\h a^\d_i \h a_j \rangle +\delta_{i,j}/2$, etc.

The BEC is initially assumed to be in a harmonic trap and we
continuously increase the strength of the optical lattice potential
until some final value, after which the potential is kept constant,
$V(x,t)=V_h(x)+s(t) E_r \sin^2{(\pi x/d)}$, with $s(t)=\exp{(\kappa
t)}-1$ for $t\leq\tau$ and $E_r=\hbar^2 \pi^2/(2m d^2)$, where
$d=\lambda/2\sin(\theta/2)$ is the lattice period, obtained by two
laser beams intersecting at an angle $\theta$. For very large $s$
and close to the ground state only one mode per lattice site is
important and the system can be approximated by the BHH:
\beq
H=\s_i\big[\nu_i \h b^\d_i\h b_i-J(\h b^\d_i\h b_{i+1}+
{\rm H.c.})+\f{U}{2}(\h b_i^\d)^2\h b_i^2\big]\,,
\label{BH}
\eeq
where the summation is over the lattice sites, $J\simeq -\int dx
\eta^*_i(x){\cal L}\eta_{i+1}(x)$ is the hopping amplitude between
the nearest-neighbor sites, $U\simeq g_{1D}\int dx |\eta_j(x)|^4$,
and $\nu_j\equiv j^2 d^2 m\omega^2/2$, with $j=0$ site at the trap
center. We may approximate the Wannier functions $\eta_i$ by the
ground state harmonic oscillator wave function with the frequency
$\omega_s=2s^{1/2}E_r/\hbar$ at the lattice site minimum
\cite{JAK98}. When we compare the TWA results to the BHH, we
frequently extract the expectation values involving $\h{b}$ using
Eq.~(\ref{projection}) with $\h{b}\sim\h{a}$. We tested that using
different projections does not affect the results. For $n_i J\agt
U$, with $n_i\equiv\< \h b^\d_i \h b_i\>$, the system is in the
superfluid regime with the long-range phase coherence and is
expected to undergo the MI transition at $n_i J\sim U$ \cite{FIS89},
resulting in a highly number squeezed ground state.

In the numerical studies of loading the BEC into an optical lattice,
we first solve the BEC ground state $\psi_0$ by evolving the GPE in
imaginary time in the harmonic trap and then diagonalize
Eq.~(\ref{Bogo}) to obtain the quasiparticle mode functions
$u_j,v_j$ and energies $\ej$. The time evolution of the ensemble of
Wigner distributed wavefunctions [Eq.~(\ref{GP})] is unraveled into
stochastic trajectories, where the initial state of each realization
for $\psi_W$ is generated according to Eq.~(\ref{field}) with the
operators replaced by the Gaussian-distributed random variables
$(\alpha_j,\alpha_j^*)$. We integrate Eq.~(\ref{GP}) using the
split-step method and in several cases the sufficient convergence is
obtained after 600 realizations. The convergence is generally slower
at higher temperatures. Unlike the 3D TWA \cite{SIN02}, the 1D
simulations do not similarly depend on the total number of
quasiparticle modes and we found the calculated results to be
unchanged when we increased the number of modes.

For the typical nonlinearity $N_0g_{1D}=100\hbar\omega l$, with
$l\equiv (\hbar/m\omega)^{1/2}$, the initial harmonically trapped
BEC is well described by the GPE with the Poisson density
fluctuations and the ratio between the interaction and the kinetic
energy $\gamma=mg_{1D}/(\hbar^2n_{1D})\alt 10^{-3}$ \cite{KHE03},
where $n_{1D}$ is the 1D atom density. The corresponding initial
Thomas-Fermi radius $R/l=(3N_0g_{1D}/2\hbar\omega
l)^{1/3}\simeq5.3$. We take $d=\pi l/8$, resulting in
$E_r=32\hbar\omega$. Within $2R$, we then have 30-35 lattice sites.
A similar number of sites has also been realized in recent
experiments in a cigar-shaped trap with $d\simeq 2.7\mu$m
\cite{HAD04}. In order to characterize the phase coherence along the
lattice, we introduce the normalized first-order correlation
function between the central well and its $i$th neighbor as
$C_i\equiv |\langle \h a^\d_0\h a_i\rangle | /\sqrt{n_0n_i}$.
\begin{figure}
\includegraphics[width=0.43\columnwidth]{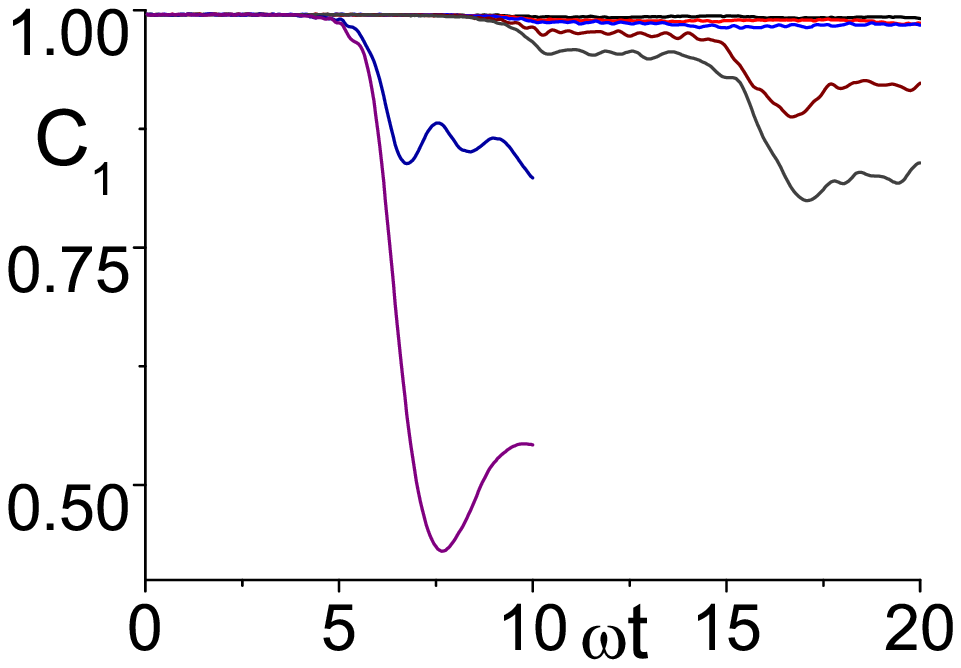}
\includegraphics[width=0.43\columnwidth]{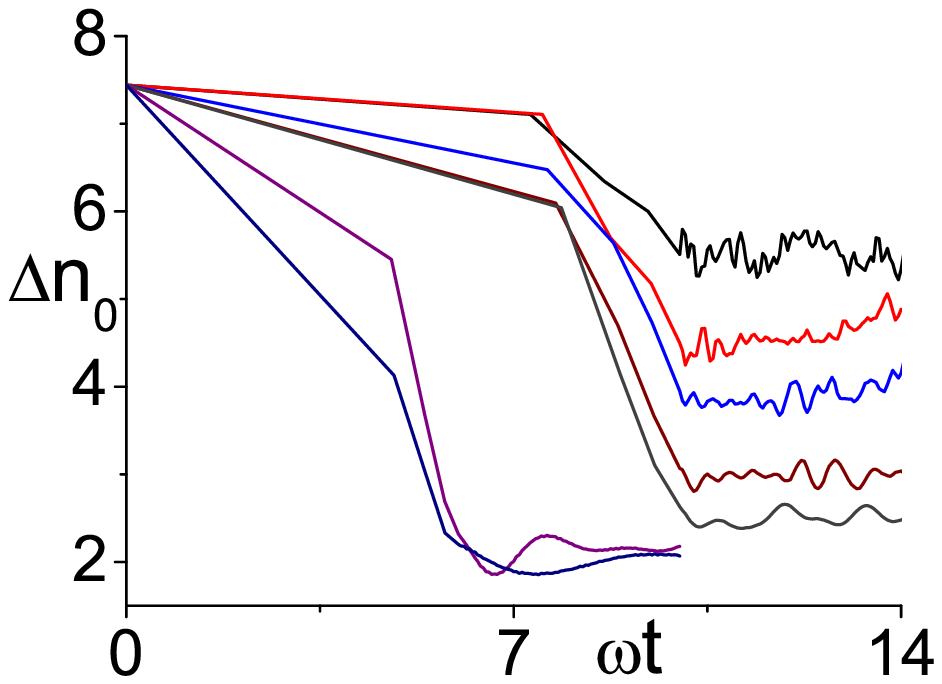}
\vspace{-0.6cm} \caption{The phase coherence between the central
well and its nearest neighbor $C_1$ as a function of time (left) at
$T=0$ for different final heights of the optical lattice (curves
from top represent $s=5,8,10,15,20,30,40$ with the atom number in
the central well $n_0\simeq$90-100). The ramping time
$\omega\tau=10$, except for $s=30,40$, $\omega\tau=6$. The number
fluctuations $\Delta n_0$ for the central well (right) for the same
runs. Here $g_{1D}=0.05\hbar\omega l$ and $N_0=2000$. The number
squeezing can be accurately fitted according to $(\Delta
n_0)^2/n_0\simeq 0.03+0.5e^{-s/8}\simeq0.03+0.3J^{0.4}$. }
\label{fig1}
\end{figure}

In Fig.~\ref{fig1} we show $C_1$ and the number fluctuations $\Delta
n_i= [\<(\h{a}_i^\d \h{a}_i)^2\>- \<\h{a}_i^\d \h{a}_i\>^2]^{1/2}$
in the central well for different final heights of the periodic
potential at $T=0$. For shallow lattices the phase coherence remains
high and steady, but for larger $s$ it is reduced and becomes
strongly oscillatory. Due to the large occupation numbers, $\Delta
n_0$ are strongly sub-Poissonian, approaching the asymptotic value
$(\Delta n_0)^2/n_0\simeq 0.03\ll 1$ for large $s$. Here the MI
transition for the ground state is expected to occur at $s\simeq
30$. However, we find $\Delta n_0\agt 1$ for all $s$, which can be
understood by the nonadiabatic loading.

For an adiabatic turning up of the lattice and for the system to
remain in its ground state, we require that the rate of change in
the tunneling amplitude to be slower than any characteristic time
scale of the system. At low lattice heights it is more difficult to
avoid exciting higher vibrational levels within one potential well,
resulting in excitations in the higher energy bands. Moreover, the
phonon mode energies $\omega_n$ in the lowest energy band decrease
with increasing lattice strength \cite{JAV99,BUR02} and for high
lattices it is more difficult to maintain the adiabaticity with
respect to these excitations. In Fig.~\ref{fig1} we find the number
squeezing to saturate around $s$=20-30, indicating the point when an
increasing number of phonon modes is excited and the loading becomes
strongly nonadiabatic. Consequently, the $s\ge 15$ cases exhibit
significant excess number fluctuations as compared to the ground
state. After a short time period over which $C_1$ remains constant,
the large $\Delta n_i$ evolve into large phase fluctuations and
$C_1$ becomes oscillatory and collapses.
\begin{figure}
\includegraphics[width=0.43\columnwidth]{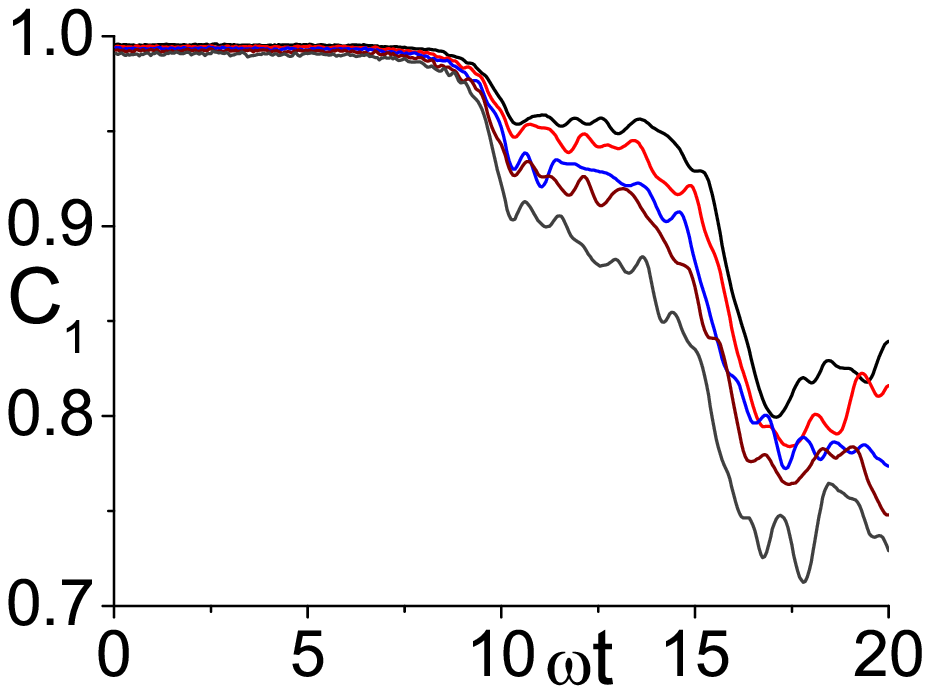}
\includegraphics[width=0.43\columnwidth]{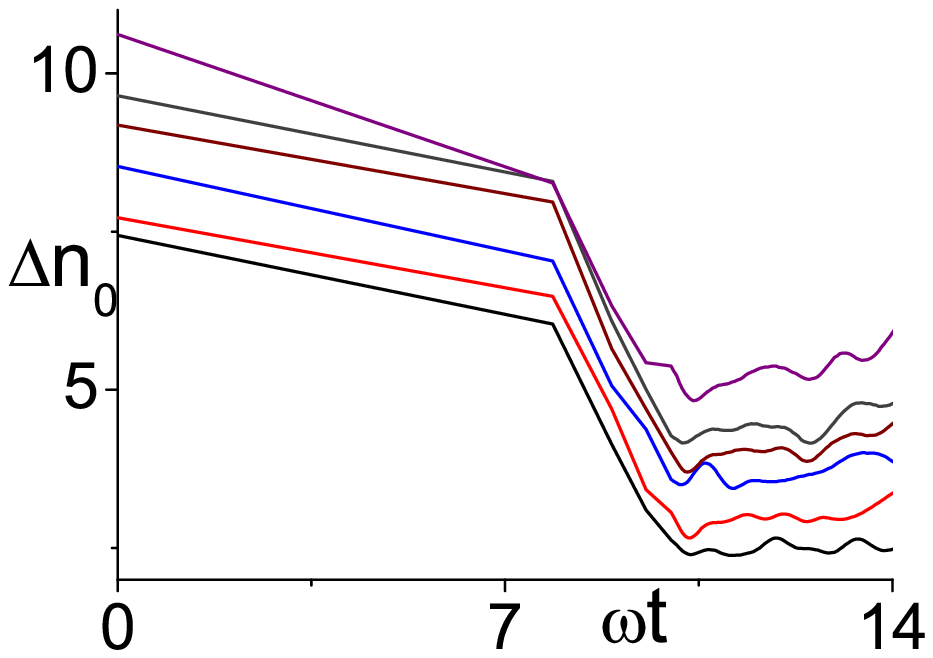}\vspace{-0.5cm}
\includegraphics[width=0.43\columnwidth]{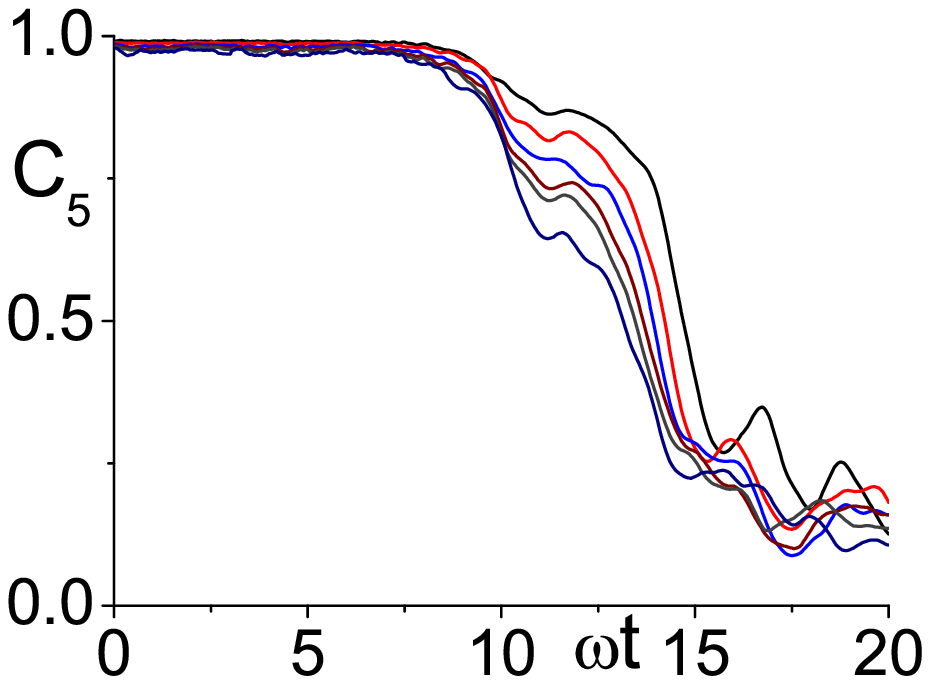}
\includegraphics[width=0.43\columnwidth]{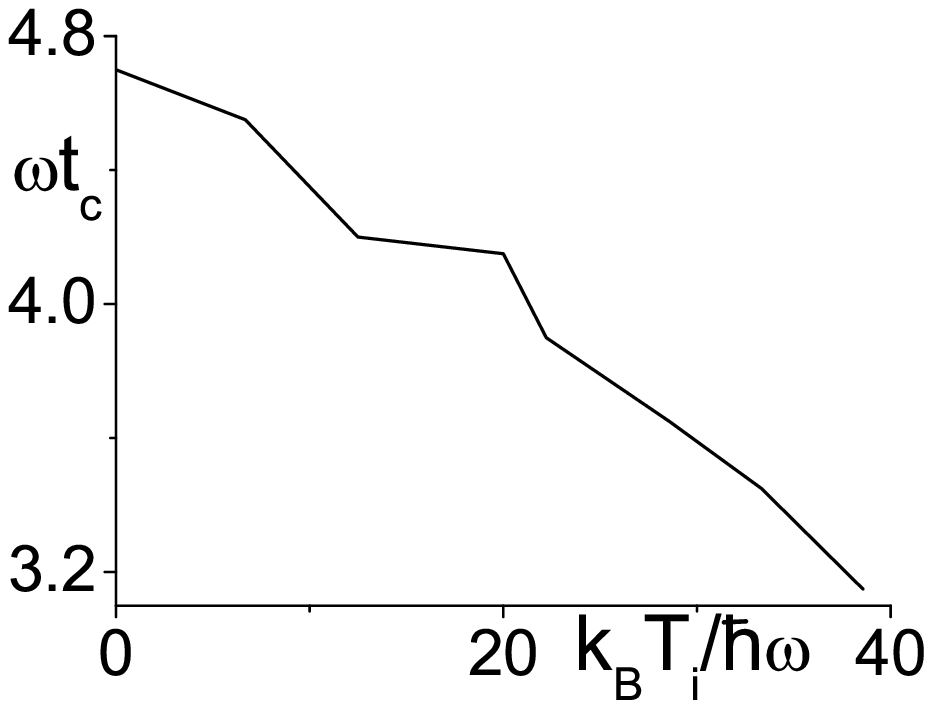}
\vspace{-0.6cm} \caption{The coherences $C_1$ (top left), $C_5$
(bottom left) and the number fluctuations $\Delta n_0$ (top right)
for initial temperatures (curves from top)
$k_BT_i/\hbar\omega=0,12.5,22.2,33.3,38.5$ ($C_5$ also with 28.5).
The phase collapse time $t_c$ (bottom right) is evaluated at
$C_5=0.5$. The same system as Fig.~\ref{fig1} with $s=20$.}
\label{fig2}
\end{figure}

The saturation of the number squeezing for strong lattices was
experimentally observed in \cite{ORZ01} for a 3D vapor in a 1D
lattice. Such a system is not tightly elongated, but we can still
make qualitative comparisons to the experimental data. Although the
saturation was assumed in \cite{ORZ01} to be an artifact of the
analysis method of the interference measurement, we also numerically
find the same saturation effect which can be explained by the
nonadiabaticity of the loading process. If the loading is
sufficiently rapid or the final lattice sufficiently high, so that
the adiabaticity breaks down for a large number of modes, the
optimal number squeezing is proportional to the ramping speed itself
and the nonlinearity. Both in Fig.~\ref{fig1} and in \cite{ORZ01}
the squeezing saturates at about 15dB when $n_iU/J\sim10^4$. The
ramping time $\tau\simeq 4000\hbar/E_r$ in \cite{ORZ01} is one order
of magnitude longer than in Fig.~\ref{fig1}, but this is compensated
by the weaker hopping amplitude $J$, so that the saturation roughly
occurs at the same value of $\omega_n\tau$.

Although the BHH (\ref{BH}) is only valid for weakly excited high
lattices, it is interesting to compare the TWA results to the
Bogoliubov approximation to the BHH. These were calculated in the
homogeneous lattice ($\nu_i=0$) in \cite{JAV99,BUR02,REY03}.
Similarly, we may diagonalize the linearized fluctuations in
Eq.~(\ref{BH}) around the ground state atom density with the
fluctuation part $\delta \h b_j = \sum_n [f_n(jd)
\h\chi_n-h_{n}^*(jd)\h\chi_n^\d]$, resulting in the number
fluctuations in each site, $(\Delta n_i)^2= n_i \sum_j |w_j|^2
(2\bar{n}_j+1)$, and the phase fluctuations between the $k$ and $l$
sites, $(\Delta\varphi_{kl})^2\equiv \<(\h\varphi_k-\h\varphi_l)^2\>
= 1/4\sum_j |r_j(kd)/\sqrt{n_k}-r_j(ld)/\sqrt{n_l}|^2
(2\bar{n}_j+1)$, where $\h n_i= \sqrt{n_i} \sum_j (w_j \h\chi_j+
w^*_j\h\chi^\d_j)$ and $\h \varphi_i= -i/(2 \sqrt{n_i}) \sum_j
(r_j\h\chi_j - r_j^*\h\chi^\d_j)$ are the number and phase
operators, with $w_j\equiv f_j-h_j$, $r_j\equiv f_j+h_j$, and
$\bar{n}_j=\<\h\chi^\d_j\h\chi_j\>$. In the homogeneous lattice with
$n$ atoms per site we have
$(\hbar\omega_q)^2=4J\sin^2{(qd/2)}[4J\sin^2{(qd/2)}+2nU]$, where
$q$ is the quasiparticle momentum \cite{JAV99,BUR02}. Moreover, for
$nU\gg J$ and $N_p$ lattice sites, $(\Delta n_i)^2\simeq \sum_q
\hbar\omega_q/(2UN_p) (2\bar{n}_q+1)$ and $(\Delta
\varphi_{k,k+1})^2\simeq \sum_q \hbar\omega_q/(4nJN_p)
(2\bar{n}_q+1)$, which at $T=0$ approximately yield
$(8nJ/U)^{1/2}/\pi$ and $(2U/nJ)^{1/2}/\pi$, respectively.
Numerically, we find the Bogoliubov results in the harmonic trap for
$\Delta n_0$ to be slightly larger and for $\Delta \varphi_{01}$
smaller than the homogeneous result. The TWA results for $\Delta
n_i$ in Fig.~\ref{fig1} are clearly larger than the ideal Bogoliubov
limit, however, $(\Delta n_0)^2/n_0\propto J^{0.4}$ still
qualitatively similar to the Bogoliubov result ($n_0U$ depends only
weakly on $s$). As argued in \cite{JAV99}, if the adiabaticity of a
phonon mode breaks down, the number fluctuations of the mode freeze
to the value that prevails at the time this occurs, i.e., when
$\omega_j\sim \zeta(t)\equiv|\partial_t J(t)/J(t)|$. Using the
homogeneous lattice result at $T=0$ we obtain $(\Delta n_i)^2\simeq
\sum_j \hbar\zeta_j(t_j)/(2UN_p)$. Since for all $j$, $\zeta_j(t_j)$
is here roughly of the order of $\omega$, we have the asymptotic
value for $s\rightarrow\infty$, $(\Delta n_i)^2\sim \hbar\omega/U$,
qualitatively similar to Fig.~\ref{fig1}. In order to study the
effect of the nonlinearity we also varied in the simulations
$N_0g_{1D}/\hbar\omega l$ from 100 to 400 for $s=20$ and found
$(\Delta n_0)^2/\sqrt{n_0}\propto U^{c}$, with $c\simeq-0.26$, as
compared to the Bogoliubov result $c=-1/2$.

In Fig.~\ref{fig2} we show $\Delta n_0$ and the coherence $C_1$ for
different initial temperatures $T_i$ for $s(\tau)=20$. Here $(\Delta
n_0)^2$ increases exponentially as a function of $T_i$. The phase
coherence $C_5$ between the central well and its 5th neighbor decays
significantly faster than $C_1$. The dependence of the phase
collapse time $t_c$ on $T_i$ is approximately linear. At $s=20$ the
effects of the harmonic trap are already significant, since the
variation of the trapping potential over five sites exceeds the
tunneling energy $\nu_5\simeq2\hbar\omega\agt n_0J$.

\begin{figure}[!t]
\includegraphics[width=0.43\columnwidth]{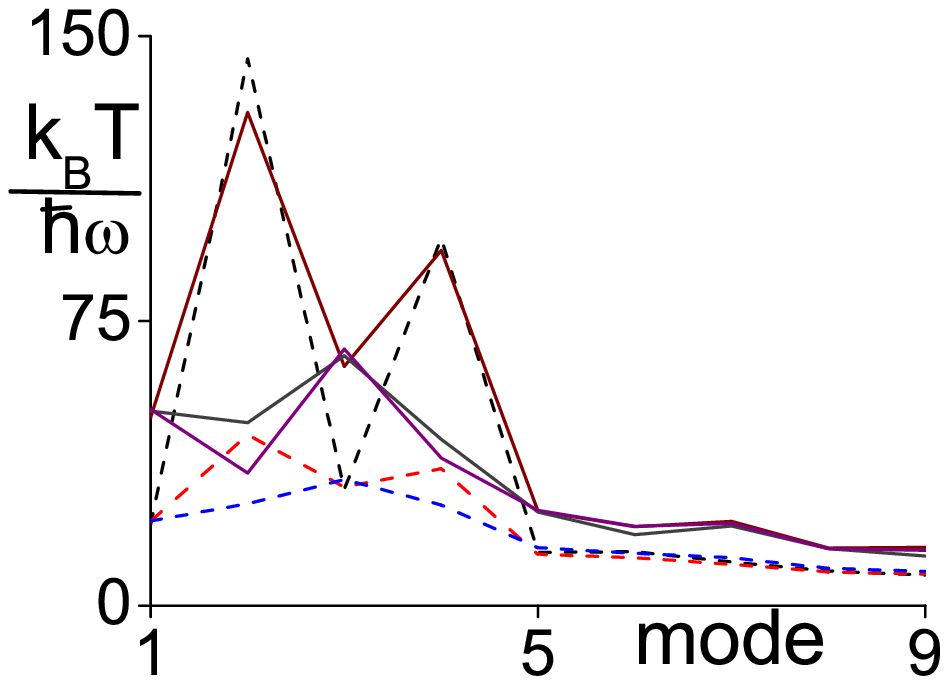}
\includegraphics[width=0.43\columnwidth]{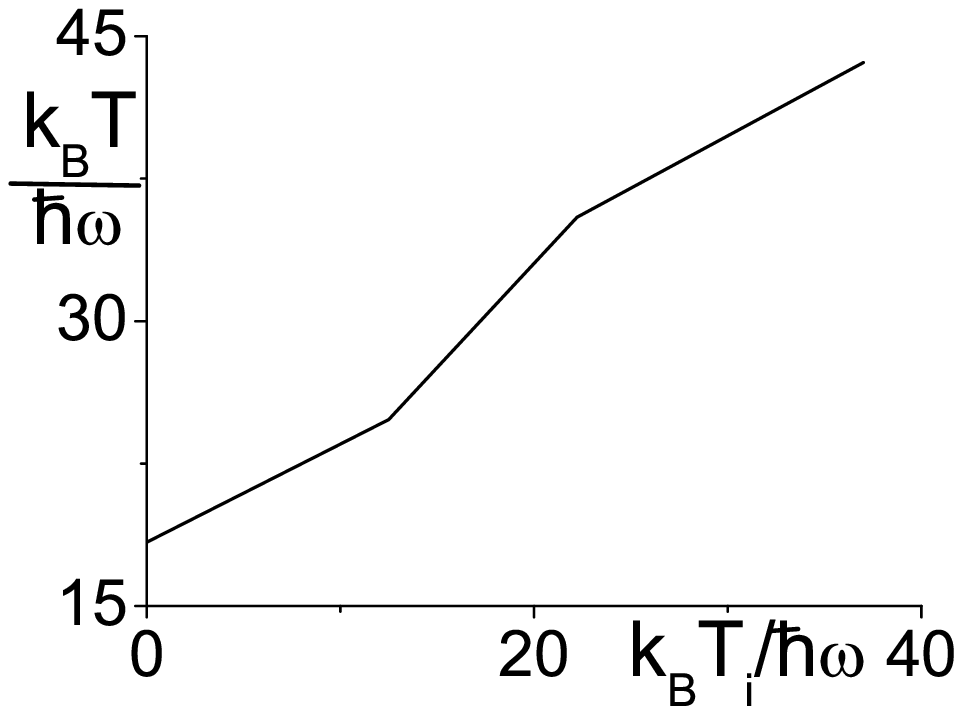}
\vspace{-0.6cm} \caption{The contribution of the lowest modes to
temperature (left) for $k_BT_i/\hbar\omega=12.5$ (dashed line) and
37 (solid line). The curves from top $\omega\tau=10,20,30$ for both
cases. The average temperature of the first five modes after the
ramping $\omega\tau=30$ (right). Here $g_{1D}=0.015\hbar\omega l$,
$N_0=2000$, and $s=5$.} \label{fig3}
\end{figure}

If the lattice potential is turned up adiabatically, the population
of each mode remains constant and temperature $T$ can change
dramatically, as the contribution of each mode to $T$ changes by the
ratio of the final and initial mode energies
$\omega_j^{(f)}/\omega_j^{(i)}$. An adiabatic increase in the
lattice strength may both increase or lower $T$, depending on
whether the excited band is occupied \cite{BLA04} and in the
experiments the condensation temperature has been found to be
sensitive to the lattice height \cite{BURG02}. In Fig.~\ref{fig3} we
estimated the population and the `temperature' of the lowest phonon
modes in the TWA simulations by evaluating the projection of
$\psi_W$ to the Bogoliubov modes of the BHH (\ref{BH}). The averages
are taken over a time period before any significant rethermalization
occurs after the ramping. The modes 2 and 4 are highly excited for
the case of short $\tau$, due to the nonadiabatic loading. The
excitations are damped out at higher $T_i$ and for $\omega\tau=30$,
corresponding to $\omega_{2,4}\tau\gg1$. It is interesting to note
that the excitations of the forth mode are only damped out when the
rate of change in the tunneling amplitude $\zeta$ is much smaller
than the corresponding mode energy, or when
$\omega_4\simeq26\zeta(\tau)$. This is more restrictive condition
than the one found in \cite{JAV99}. For $\omega\tau\alt3$, the
variation of $T_i$ is already completely dominated by the
excitations due to the rapid turning up of the lattice.

The advantage of 1D lattices is that the lattice spacing can be
easily modified by using non-parallel lasers. A large spacing could
even allow the scattering of light, or the Bragg spectroscopy, from
individual lattice sites and the separate optical detection of
number fluctuations in each site, using a similar analysis to
\cite{JAV03}. Moreover, an interference measurement on the expanding
atoms can provide detailed information about the coherence
\cite{HAD04}.

We studied the loading of a harmonically trapped BEC into an optical
lattice. In a good agreement with experiments \cite{ORZ01}, we found
the number squeezing to saturate for high lattices, which can be
explained by the finite ramping time of the lattice potential. It is
numerically more demanding to study a truly adiabatic loading for
strong lattices. However, it would be particularly interesting to
examine the validity of the TWA close to the MI ground state. Our
analysis seems to indicate that, in the case of lattices with large
filling factors, the ramping time required to reach the MI state may
be very long and can be demanding in actual experiments. In the
lattice experiments the atoms are also coupled to environment,
resulting in dissipation with the system relaxing towards its ground
state. We could improve our model, e.g., by incorporating the
spontaneous emission due to the lattice lasers. This would introduce
also a dynamical noise term in Eq.~(\ref{GP}). However,
experimentally spontaneous emission can also be avoided since, e.g.,
with intense CO$_2$ lasers the spontaneous emission rate is very low
\cite{OHA99}. Finally, our TWA studies could also be extended to the
finite temperature damping of nonequilibrium oscillations in a
multi-well BEC what has previously been studied in double-well BECs
\cite{RUO98b}.

\acknowledgments{We acknowledge financial support from the EPSRC.}

\end{document}